\documentclass[twocolumn,english,superscriptaddress,notitlepage]{revtex4-1}
\usepackage[T1]{fontenc}
\usepackage[latin9]{inputenc}
\usepackage{graphicx}

\makeatletter
\usepackage{babel}

\makeatother

\usepackage{babel}
\begin{document}

\title{Scalar chiral spin-1/2 order on kagome lattices in Nd$_3$Sb$_3$Mg$_2$O$_{14}$}

\author{A. Scheie}

\address{Institute for Quantum Matter and Department of Physics and Astronomy, Johns Hopkins University, Baltimore, MD 21218}

\author{M. Sanders}

\address{Department of Chemistry, Princeton University, Princeton, NJ
08544 }

\author{J. Krizan}

\address{Department of Chemistry, Princeton University, Princeton, NJ
08544 }

\author{Yiming Qiu}
\address{NIST Center for Neutron Research, National Institute of Standards and Technology, Gaithersburg, MD 20899}

\author{R.J. Cava}

\address{Department of Chemistry, Princeton University, Princeton, NJ
08544 }

\author{C. Broholm}

\address{Institute for Quantum Matter and Department of Physics and Astronomy, Johns Hopkins University, Baltimore, MD 21218}
\address{NIST Center for Neutron Research, National Institute of Standards and Technology, Gaithersburg, MD 20899}
\address{Department of Materials Science and Engineering, Johns Hopkins University, Baltimore, MD 21218}

\date{\today}

\begin{abstract}
We introduce $\mathrm{Nd_{3}Sb_{3}Mg_{2}O_{14}}$ with ideal kagome lattices of  neodymium ions in ABC stacking. Thermodynamic measurements show a Curie-Weiss temperature of $\Theta_{CW}=-0.12$~K,  a Nd$^{3+}$ spin-1/2 Kramers doublet ground state, and a second order phase transition at $T_N=0.56(2)$~K. Neutron scattering reveals non-coplanar scalar chiral $\bf k =0$ magnetic order with a correlation length exceeding 400 \AA = 55 $a$ and an ordered moment of 1.79(5)~$\mu_B$. This order includes a canted ferromagnetic component perpendicular to the kagome planes favored by  Dzyaloshinskii-Moriya interactions. 
\end{abstract}
\maketitle
Composed of corner-sharing triangles that frustrate conventional magnetic order, the kagome lattice supports exotic forms of  magnetism. The Ising kagome model is paramagnetic,\cite{syozi} the quantum spin-1/2 model forms a quantum spin liquid,\cite{huse,punk, simeng,balents,essafi} while fluctuation induced order is expected for the classical Heisenberg model.\cite{chubukov,richey} While there may not be an "ideal" experimental realization of any of these particular models, each kagome related magnet presents opportunities to explore the effects of distinct interactions on an underlying massively degenerate manifold.\cite{takano,obradors,ramirez,hagemann, hiroi,shores,helton,okamoto} Here we report the discovery of an effective spin-1/2 kagome magnet $\mathrm{Nd_{3}Sb_{3}Mg_{2}O_{14}}$ with scalar chiral spin order that supports topologicaly protected quantum transport.\cite{ohgushi}

 Just like the celebrated herbertsmithite kagome system,\cite{shores,helton} $\mathrm{Nd_{3}Sb_{3}Mg_{2}O_{14}}$  is based on selective non-magnetic doping of the (Fd$\bar{3}$m) pyrochlore lattice of corner-sharing tetrahedra.\cite{Gardner}  The  parent compound is  the disordered pyrochlore $\rm Nd_2SbMgO_7$, which features an ordered (Nd$^{3+}$) and a disordered (Mg$^{3+}$ and Sb$^{5+}$) sublattice of corner-sharing tetrahedra. But while the nominally non-magnetic zinc sublattice of herbersmithite contains ~15\% magnetic copper ions, the different charge and size of Sb$^{5+}$ ($r\approx 74$~pm) and Nd$^{3+}$ ($r\approx 125$~pm)  lead to an \textit{ordered}  layered rhombohedral structure for $\mathrm{Nd_{3}Sb_{3}Mg_{2}O_{14}}$ with  Nd$^{3+}$ on a crystallographically ideal kagome lattice and  Mg$^{3+}$ and Sb$^{5+}$ cations segregated onto fully occupied non-magnetic triangular lattices. These intervening layers should reduce magnetic interactions between kagome planes.\cite{Fu,Li,Zouari} The magnetism is based on Nd$^{3+}$ where spin orbit coupling prevails over crystal fields to produce a Kramers doublet protected by time reversal symmetry. Thus we have in  $\mathrm{Nd_{3}Sb_{3}Mg_{2}O_{14}}$ a rare opportunity to study spin-1/2 kagome magnetism without significant quenched disorder. 

 $\mathrm{Nd_{3}Sb_{3}Mg_{2}O_{14}}$ is isostructural to three recently reported compounds, $\mathrm{Er_{3}Sb_{3}Mg_{2}O_{14}}$, $\mathrm{Dy_{3}Sb_{3}Mg_{2}O_{14}}$, and $\mathrm{Gd_{3}Sb_{3}Mg_{2}O_{14}}$,\cite{dun} which also form crystallographically ideal ordered kagome lattices. An isostructural series based on Zn in place of Mg was also recently communicated.\cite{Sanders} This paper reports the first determination of a magnetic ordered structure in this family of materials. The jarosites also form ideal ordered kagome latices organized in ABC stacking.\cite{takano} Their magnetism arises from the semi-classical spin of transition metal ions and is characterized by long range order that maintains the kagome unit cell.\cite{inami,lee,ballou} The langasites are rare earth based magnets with "breathing" kagome lattices where the triangular magnetic units come in at least two variants with longer range interactions that lead to incommensurate magnetism.\cite{Bordet,Simonet}  
$\mathrm{Nd_{3}Sb_{3}Mg_{2}O_{14}}$ appears to form an ideal kagome lattice of effective quantum spins-1/2 with anisotropic interactions.

Polycrystalline $\mathrm{Nd_{3}Sb_{3}Mg_{2}O_{14}}$ was synthesized from a stoichiometric mixture of $\mathrm{Nd_{2}O_{3}}$, $\mathrm{Mg(OH)_{2}}$, and $\mathrm{Sb_{2}O_{3}}$ using standard solid state methods. To determine the crystal structure, high resolution x-ray powder diffraction data were acquired at room temperature using beamline 11-BM at the Advanced Photon Source (APS) at Argonne National Laboratory ($\lambda=0.414183\mathrm{\mathring{A}}$) and analyzed through the Rietveld method as implemented in Fullprof.\cite{FullProf}
The magnetic susceptibility of $\mathrm{Nd_{3}Sb_{3}Mg_{2}O_{14}}$ was measured between 1.8 and 300 K in a Quantum Design Physical Properties Measurement System (PPMS) in an applied field of 0.5 T.\cite{disclaimer} The magnetization was found to be proportional to magnetic field for all temperatures above 1.8 K and fields less than 1 T. Therefore M/H at $\mu_0 H=0.5$~T was used as a measure of the susceptibility. The specific heat $C(T)$ was measured between 0.1~K and 300~K in zero applied magnetic field in a PPMS equipped with a Quantum Design dilution refrigerator.\cite{disclaimer} The sample was a dense polycrystalline pellet with a mass of 0.72 mg.

To investigate the magnetic order of $\mathrm{Nd_{3}Sb_{3}Mg_{2}O_{14}}$,
neutron diffraction measurements were carried out on MACS at the NCNR.\cite{rodriguez} The monochromator was set for vertical focusing of 5 meV neutrons and the horizontal beam divergence was defined by a 50 mm wide beam aperture before the monochromator and the 20~mm sample diameter to be $1.5$\textdegree. 
10.2 g of loose $\mathrm{Nd_{3}Sb_{3}Mg_{2}O_{14}}$ powder was loaded into an aluminum can under 10 bar of helium gas at room temperature. The can was closed with an indium sealed copper lid and cooled to the 50 mK base temperature of a KelvinOx-100 dilution fridge with 100~$\mu$W cooling power at 100 mK.\cite{disclaimer} A 0.02 Tesla magnetic field was applied throughout to maintain aluminum in the normal state. We acquired  diffraction data for 24 hours at base temperature ($T<0.1$~K) and for 24 hours at $T=0.8$~K. The neutron diffraction data were analyzed using the FullProf \cite{FullProf} and SARAh \cite{Wills} programs.

\begin{figure}
\centering\includegraphics[scale=0.6]{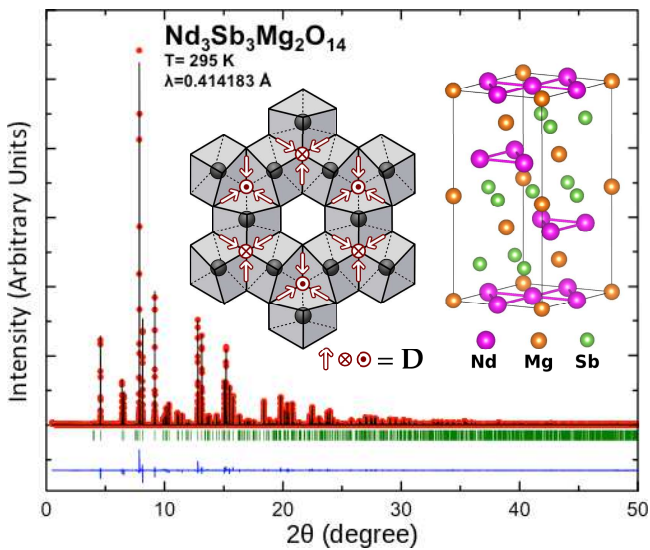}

\caption{Rietveld refinement of 11-BM x-ray diffraction data from $\mathrm{Nd_{3}Sb_{3}Mg_{2}O_{14}}$ at $T=295$~K,
showing the experimental data points (red), the Rietveld profile  (black),
and their difference  (blue). The green marks indicate Bragg reflections.
Also displayed are the stacked kagome planes of Nd atoms in $\mathrm{Nd_{3}Sb_{3}Mg_{2}O_{14}}$ and the nearest neighbor coordination relevant to spin-anisotropy, superexchange, and Dzyaloshinskii-Moriya (DM) interactions within  kagome planes. The oxygen atoms around each Nd site form scalenohedrons that tilt in alternating directions allowing for the indicated directions of $\bf D$ that define the DM interactions: $\bf D\cdot (J_i\times J_j)$. }

\label{XrayRefinement}
\end{figure}

Figure \ref{XrayRefinement} shows a high quality fit of the x-ray diffraction pattern of $\mathrm{Nd_{3}Sb_{3}Mg_{2}O_{14}}$
to the layered Kagome crystal structure with an ideal Kagome plane of Nd and  non-magnetic neighboring planes. The rhombohedral lattice parameters of $\mathrm{Nd_{3}Sb_{3}Mg_{2}O_{14}}$ ($a=7.422(9) \mathrm{\mathring{A}}$ and $c=17.502(4) \mathrm{\mathring{A}}$) are closely related to those of the cubic pyrochlore $\mathrm{Nd_{2}ScNbO_{7}}$ ($a_{p}=10.5336(3)\mathrm{\mathring{A}}$): by $a=b\approx a_{p}/\sqrt{2}$ and $c\approx\sqrt{3}a_{p}$.\cite{Zouari} The broken symmetries associated with the cubic to rhombohedral transision reflects the loss of intersecting Kagome planes in the pyrochlore lattice due to the planar ordering of Nd$^{3+}$, Sb$^{5+}$, and Mg$^{2+}$.
In the Kagome compound, Nd is found in an 8-coordinated distorted cube (a "scalenohedron") of O atoms with Nd-O bond distances ranging from 2.342(3)$\mathring{\mathrm{A}}$
to 2.580(4)$\mathring{\mathrm{A}}$. 

\begin{figure}
\centering\includegraphics[scale=0.45]{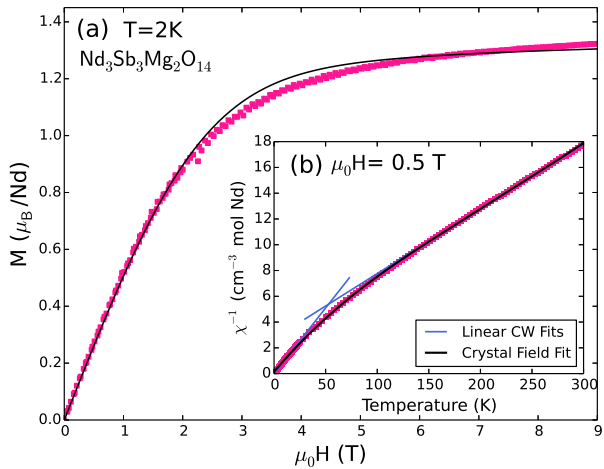}

\caption{Magnetization data for a powder sample of $\mathrm{\mathrm{Nd_{3}Sb_{3}Mg_{2}O_{14}}}$. (a) Magnetization as a function of applied field. (b) Inverse susceptibility $H/M$ in an applied field of 0.5 T.  The dashed lines are linear fits to distinct $T-$ranges. For $1.8 {\rm K}<T<10{\rm K}$ we find an effective moment of 2.50(5)~$\mu_B$ and a Curie-Weiss temperature $\Theta_{CW}=-0.12(1)$~K. The higher temperature fit for $150 {\rm K}<T<300{\rm K}$ yields the full effective moment of 3.56(5)~$\mu_B$ and a large negative intercept of $-54.5(5)$~K, which is indicative of the crystal field energy scale. The solid lines are a single consistent fit to the data in (a) and (b) based on a mean field model described in the text.}

\label{Magnetization} \label{Susceptibility} 
\end{figure}

Our first aim is to establish the nature of the spin hamiltonian in $\mathrm{Nd_{3}Sb_{3}Mg_{2}O_{14}}$ through high temperature and high field measurements. Fig.~\ref{Susceptibility}(b) shows the temperature dependence of the inverse susceptibility. Curie-Weiss fits in the low $T$ and high $T$ ranges  yield effective moments of $\mu_{eff}(1.8 {\rm K}<T<10{\rm K})=2.50(5)~\mu_B$ and $\mu_{eff}(150 {\rm K}<T<300{\rm K})=3.56(5)~\mu_B$. The latter value is similar to the free ion value of $g_J\sqrt{J(J+1)}\mu_B=3.62 \mu_{B}$ for the $J=9/2$ multiplet of Nd$^{3+}$. The ratio between the high and low $T$ effective moments is 0.70(2), which is consistent with the value of $1/\sqrt{2}\approx 0.707$ for an easy-plane ion in the large $J$ limit with a random orientation distribution. Similar low $T$ effective moments have been observed for several other easy plane neodymium pyrochlores oxides, such as $\mathrm{Nd_{2}ScSbO_{7}}$ (2.58 $\mu_{B}$), $\mathrm{Nd_{2}Pb_{2}O_{7}}$ (2.55$\mu_{B}$), $\mathrm{Nd_{2}Zr_{2}O_{7}}$
(2.54$\mu_{B}$), and $\mathrm{Nd_{2}Sn_{2}O}$ (2.63 $\mu_{B}$).\cite{Hallas,Strobel,Hatnean,Matsuhira} 
The energy scale (Curie-Weiss temperature) extracted from the high $T$ fit $\Theta (150 {\rm K}<T<300{\rm K})=-54.5(5)$~K is indicative of the overall crystal-field level splitting for $\mathrm{Nd_{3}Sb_{3}Mg_{2}O_{14}}$ rather than the inter-site exchange energy. 
From the low $T$ Curie Weiss fit, we extract an estimate for the sum of all exchange constants associated with each Nd site of  $\Theta (1.8 {\rm K}<T<10{\rm K})=-0.12(1)$~K. The large ratio  $\Theta (150 {\rm K}<T<300{\rm K})/\Theta (1.8 {\rm K}<T<10{\rm K})=454(5)$ indicates the lowest lying easy plane doublet $|J,m_J\rangle=|9/2,\pm 1/2\rangle$ should suffice to describe the collective magnetism of $\mathrm{Nd_{3}Sb_{3}Mg_{2}O_{14}}$. 

Shown in Fig.~\ref{Magnetization}(a) is the field dependent magnetization, $M(H)$, at 2 K measured for a loose powder. The slight hysteresis in the magnetization data that is particularly apparent near $\mu_0H=2$~T may be associated with some reorientation of powder grains that places easy planes parallel to the applied field direction. Nonetheless the measured saturation magnetization is close to the value of  $M_{sat}=g_J(11/6)\mu_B=1.33~\mu_B$ corresponding to random orientation of the easy plane $|9/2,\pm 1/2\rangle$ doublet ground state with respect to the applied field and a Zeeman energy that is far less than the first excited crystal field level.

To quantify anisotropy and exchange parameters, a consistent fit was carried out to $\chi(T)$ and $M(H)$ based upon the following mean field Hamiltonian: $\mathcal{H}=D_{\rm CF} J_{z}^{2} + g_J \mu_{B}\mu_0 {\bf H}_{\rm eff}\cdot {\bf J}$.  Here $D_{\rm CF}>0$ represents the crystal field easy plane anisotropy energy, and $g_J$ is the Land\'{e} factor. The effective field is ${\bf H}_{\rm eff}=(H+\lambda M){\hat {\bf H}}$ where ${\bf H}=H{\hat {\bf H}}$ is the applied field, $M$  the induced  magnetization per volume unit,  $\lambda=(V/N)(g_j\mu_B)^{-2}{\cal J}_{tot}$, and ${\cal J}_{tot}=(1/2N)\Sigma_{ij}{\cal J}_{ij}$ is the  total exchange energy per spin. The corresponding fitted parameters are $D_{\rm CF}=28.5(3)\mathrm{meV}$, $g_J=0.7235(9)$, and ${\cal J}_{tot}=-23(1)\ \mu\mathrm{eV}$. The excellent and consistent description of the data (solid lines in Fig.~\ref{Susceptibility}) and consistency with $g_J=0.7273$ for the Nd$^{3+}$ ion indicate the low temperature magnetism of $\mathrm{Nd_{3}Sb_{3}Mg_{2}O_{14}}$ can be described in terms of an easy-plane Kramers doublet with net antiferromagnetic interactions.

\begin{figure}
\centering\includegraphics[scale=0.5]{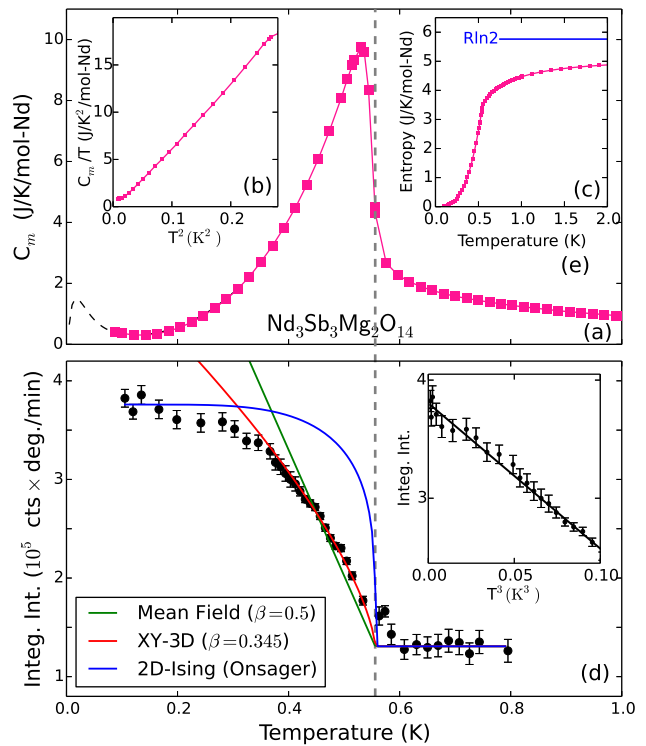}

\caption{(a) Low-temperature specific heat of $\mathrm{Nd_{3}Sb_{3}Mg_{2}O_{14}}$. The low $T$ upturn  results from hyperfine enhanced nuclear specific heat from $^{143}$Nd and $^{145}$Nd. The dashed line shows calculated nuclear contribution with a fitted ordered Nd moment of 1.5(1)~$\mu_B$. (b) $C_m/T$ versus $T^2$ after subtracting the low $T$ nuclear contribution to the specific heat. (c) Change in electronic entropy obtained through an integral of $C_m/T$. (d) Integrated intensity of the (101) peak versus $T$ along with several fits using various critical exponents  $\beta$ while fixing $T_{N} \equiv 0.56~\mathrm{K}$ as determined from the specific heat data (vertical dashed line). (e) Integrated intensity versus $T^3$ in the low $T$ regime where the reduction in sublattice magnetization is given by spin wave theory to be $\propto T^3$. Error bars represent one standard deviation.}

\label{Specific Heat} 
\end{figure}

Turning  to  collective phenomena, Figure \ref{Specific Heat}(a) shows low temperature specific heat data. Below 0.6 K, the specific heat strongly increases, with an inflection point at 0.56(2) K. The $\lambda-$like anomaly indicates a second order phase transition at $T_N=0.56(2)~$K, which we shall shortly show is to a three dimensional long range ordered canted antiferromagnetic state that carries a finite scalar chirality. 
For temperatures below 0.2 K  a slight upturn in $C(T)$ is ascribed to a hyperfine field enhanced nuclear Schottky anomaly. The dashed line in this $T-$regime was fit based on this hypothesis and the known hyperfine parameters \cite{hyperfine} for the stable spin-full isotopes of neodymium (12.18\%  $^{143}$Nd and 8.29\% $^{145}$Nd both with spin-7/2) and indicates an ordered Nd moment of 1.5(1)~$\mu_B$.

By integrating $C/T$, the $T-$dependent change in entropy (computed after subtracting the nuclear Schottky contribution to the specific heat) was inferred (Fig.~\ref{Specific Heat}(c)). While significant  entropy is liberated at the magnetic phase transition, the amount falls short of  $R\ln 2$  anticipated for a ground state doublet. Neglecting any low $T$ configurational entropy, this indicates significant spin correlations for $T>T_N$~K. Fig.~\ref{Specific Heat}(b) shows a detail of the electronic low $T$ specific heat (after subtraction of the nuclear term). The limiting $T^3$ behavior is consistent with linear dispersive spine wave excitations in three dimensions. The corresponding average spin wave velocity extracted from a low $T$ fit to $C_m(T)=(\pi^{2}/15)R(k_{B}T/\hbar v_{\rm sw})^{3}v_0$ is $v_{\rm sw}=55(1)$~m/s.  Here $v_0$ is the magnetic unit cell volume. This estimate of the spin-wave velocity  is not inconsistent with the value of $v_{\rm sw}=\sqrt{2}|{\cal J}_{tot}|Sa/\hbar=22(1)$~m/s inferred from susceptibility data via the low $T$ Curie-Weiss fit.

\begin{figure}
\centering\includegraphics[scale=0.36]{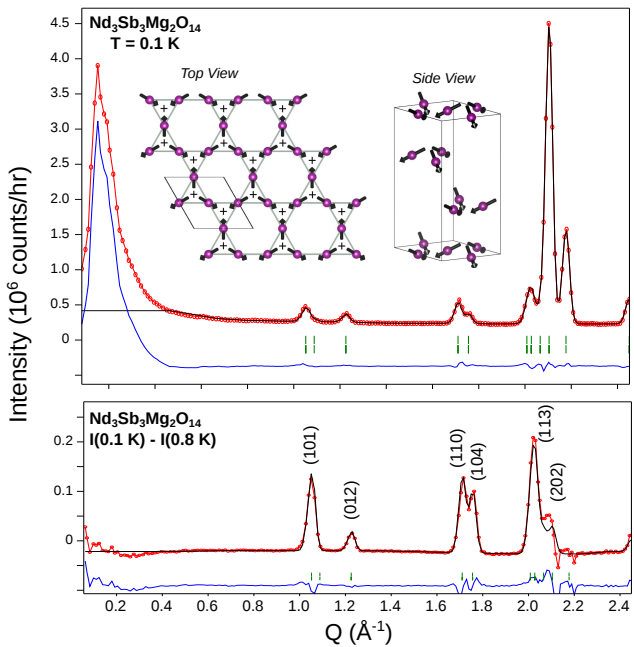}

\caption{Rietveld refined neutron diffraction data for $\mathrm{Nd_{3}Sb_{3}Mg_{2}O_{14}}$.  (a) Low $T$ data along with images of the corresponding scalar chiral umbrella-like spin structure. (b) Magnetic diffraction data at $T=0.1$~K isolated form nuclear scattering by subtracting nuclear diffraction data acquired at $T=0.8$~K$>T_N$.}
\label{flo:NeutronData}

\end{figure}

To directly probe the magnetic order parameter we acquired high statistics  neutron diffraction data at 0.1~K and 0.8~K. The temperature difference in Fig.~\ref{flo:NeutronData}(b) reveals increased intensity at the location of several nuclear Bragg peaks upon cooling below $T_N$. This is consistent with long range magnetic order with a propagation vector of $\mathbf{k}=(0,0,0)$. Fitting the excess diffraction at (101) to the convolution of a lorentzian peak with a gaussian peak that fits the high $T$ (101) peak and so accounts for the instrumental resolution, we obtain a lower bound on the magnetic correlation length at $T=0.1$~K of $\xi$ > 400\AA = 55$a$. Fig. \ref{Specific Heat}(d) shows the peak area (extracted from gaussian fits) of the (101) peak versus $T$. The onset for $T<T_N$ links the extra diffraction intensity to the squared order parameter of the phase transition. Knowing the critical temperature from the specific heat data, we compare the $T-$dependent diffraction to that anticipated for phase transitions of different universality classes. The data are inconsistent with mean field criticality and exclude a 2D Ising transition. Fitting the data within 20\% of $T_N$ yields a critical exponent $\beta=0.34(4)$ which is consistent with 3D Ising, XY, and Heisenberg criticality. Both the diffraction line-shape and the experimental value for $\beta$ however, exclude a 2D  transition. 

To determine the magnetic structure we carried out a Rietveld refinement of the $T=0.1$~K data, which contains nuclear as well as magnetic diffraction. Refinement of the nuclear component started from the high $T$ structure determined by X-ray diffraction (Fig. 1). We used representation analysis as implemented in the SARAh program to classify magnetic structures according to the three irreducible representation (IRs) of the $R\overline{3}m$ space group. Only the three dimensional representation denoted $\Gamma_3$ provides an acceptable fit (Fig.~\ref{flo:NeutronData}). For comparison the  goodness of fit parameter $\chi^2$ increases by a factor of 20 and 59 for the six dimensional $\Gamma_5$ irrep and the one dimensional $\Gamma_1$ irrep respectively. The magnetic diffraction intensity was isolated from the nuclear component by subtracting 0.8 K data from the 0.1 K data in Fig.~\ref{flo:NeutronData}(b). Apart from a slight discrepancy near (202), which results from systematic errors associated with subtracting intense nuclear peaks acquired at different temperatures, the apportionment of magnetic and nuclear diffraction associated with the fit in Fig.~\ref{flo:NeutronData}(a) is consistent with the difference data.

With moments canted along a common c-direction and pointing either ``all-into''  or ``all-out''  from their triangles, the $\Gamma_3$ magnetic structure has an umbrella-like character and carries a net moment along $c$. The  ordered magnetic moment extracted from refinement is $1.79(5)~\mu_{B}$. This is consistent with the observed saturation magnetization assuming an easy-plane moment and  with the ordered moment inferred from the nuclear Schottky anomaly. Related  umbrella-like  structures were previously reported in jarosite compounds with $S\ge 3/2$. For $\rm KFe_3(OH)_6(SO_4)_2$ the out of plane spin component alternates between layers so the unit cell is doubled,\cite{Nishiyama-Jer,Matan-Jer} while $\rm KCr_3(OH)_6(SO_4)_2$ appears to have the same structure as $\mathrm{Nd_{3}Sb_{3}Mg_{2}O_{14}}$ though  spin canting was not reported.\cite{lee} We find a spin canting angle of $\eta= 30.6(5)^{\circ}$ for $\mathrm{Nd_{3}Sb_{3}Mg_{2}O_{14}}$, which corresponds to a net ferrmagnetic moment of $0.93(3)~\mu_{B}$/Nd. This dwarfs the canting angle of $\eta\approx 2^{\circ}$ found for the jarosites.

Dzyaloshinskii-Moriya (DM) interactions are allowed by symmetry when the midpoint between magnetic ions is not a point of inversion.
In jarosites, DM interactions favor the alternating umbrella structure\cite{ballou,Elhajal-DM,Matan-Jer,Yildirim-Jer}. Likewise in $\mathrm{Nd_{3}Sb_{3}Mg_{2}O_{14}}$ oxygen atoms surrounding each Nd site are tilted in an alternating pattern that allows DM vectors in the mid-plane between interacting spins with the orientations indicated in Figure \ref{XrayRefinement}. The corresponding interactions ${\bf D}\cdot ({\bf J}_i\times{\bf J}_j)$ favor the $\Gamma_3$ uniform umbrella structure that we infer from diffraction data.\cite{Elhajal-DM,Matan-Jer,ballou} Given the strongly anisotropic nature of the $|9/2,\pm 1/2\rangle$ state the large canting angle is to be expected. For semi-classical spins $\eta$ is related to a ratio of interactions as follows\cite{Yildirim-Jer}
\begin{equation}
\sin 2\eta=|\frac{2D_{\perp}}{\sqrt{3}J-D_z}|,
\end{equation}
Here $D_{\perp}=|{\bf D}\times{\bf \hat{z}}|$, $D_z={\bf D}\cdot {\bf \hat{z}}$, and $J=J_1+J_2$ is the sum of near and next nearest neighbor exchange interactions. If we were to use this expression for quantum spins and neglect $D_z$ the expression yields  $D_{\perp}/J=0.8$ for $\mathrm{Nd_{3}Sb_{3}Mg_{2}O_{14}}$. While there is presently insufficient information to determine all relevant interactions for $\mathrm{Nd_{3}Sb_{3}Mg_{2}O_{14}}$, it is interesting to note that a classical spin model with DM interactions on the kagome lattice should undergo a finite temperature 2D phase transition.\cite{Yildirim-Jer,Elhajal-DM,rigol}. The phase transition in $\mathrm{Nd_{3}Sb_{3}Mg_{2}O_{14}}$ at $T_N=0.56(2)$~K is  clearly a 3D transition but the reduced change in entropy at $T_N$ is consistent with a quasi-2D regime for $T>T_N$ that could be a 2D quantum spin liquid.\cite{huh}

Through detailed bulk and neutron scattering experiments on $\mathrm{Nd_{3}Sb_{3}Mg_{2}O_{14}}$, we have provided a first atomic scale view of magnetism in a new family of rare earth kagome magnets. The non-coplanar $\bf k=0$ umbrella structure we observe carries uniform magnetization along $c$ and is favored by the staggered Dzyaloshinkii-Moriya interactions anticipated for rare earth ions on this lattice. The corresponding  scalar chiral spin structure implies Berry curvature, which should produce an anomalous thermal Hall effect and topologically protected edge states below and possibly even above $T_N$.\cite{Chirality}

We thank Elizabeth Seibel and Quinn Gibson for helpful discussions. We also thank Ovi Garlea and Stuart Calder for their help during early stages of the project. This work was supported through the Institute for Quantum Matter at Johns Hopkins University, by the U.S. Department of Energy, Division of Basic Energy Sciences, Grant DE-FG02-08ER46544. Scheie was supported through the Gordon and Betty Moore foundation under the EPIQS program GBMF4532. Use of the Advanced Photon Source at Argonne National Laboratory was supported by the U. S. Department of Energy, Office of Science, Office of Basic Energy Sciences, under Contract No. DE-AC02-06CH11357. Use of the NCNR facility was supported in part by the National Science Foundation under Agreement No. DMR-1508249.


\begin{thebibliography}{10}
\bibitem{disclaimer}The identification of any commercial product or trade name does not imply endorsement or recommendation by the National Institute of Standards and
Technology.
\end{thebibliography}
\end{document}